\documentclass[letterpaper,10.75pt]{article}
\usepackage[total={17cm,24cm},top={2cm},left={2.5cm}]{geometry}
\usepackage[utf8]{inputenc}
\usepackage{multicol,graphicx,anyfontsize,hyperref}
\usepackage{graphicx}
\usepackage{marginnote}
\usepackage{fancyhdr}
\usepackage[shortlabels]{enumitem}
\usepackage{amsfonts,amssymb,amsmath,wrapfig,lipsum}
\usepackage{float}
\usepackage{xcolor,cancel,comment,multirow}
\usepackage{pdfpages} 
\usepackage[font=footnotesize]{caption}
\usepackage{multicol}

\providecommand{\keywords}[1]
{\\
  \small\\
  \textbf{\textit{Key words---}} #1
}

\begin{document}

	\renewcommand{\refname}{Bibliography}
	\begin{flushleft}
		\scriptsize{ \hfill January - May 2022 Semester}
		\rule[.2cm]{\textwidth}{.001cm}
	\end{flushleft}
	\vspace{-.28cm}
	\begin{minipage}{16cm}
		\centering
		\begin{tabular}{c}\\
			\resizebox{10cm}{!}{\Huge{\sl{\textbf{Brain tumor detection using }}}}\\
			
			    \resizebox{13cm}{!}{\Huge{\sl{\textbf{artificial convolutional neural networks}}}}\\

		\end{tabular}\\
	\end{minipage}\\\begin{minipage}{3cm}
	
	\end{minipage}\\
	\rule{\textwidth}{.1cm}
	\begin{center}
		\normalsize{\textbf{J. Melchor, B. Sotelo, J. Vera, H. Corral, M. I. J. Ramírez}}\\
			
				\normalsize{Circuito Universitario S/N, Nuevo Campus Universitario C.P. 31125, Chihuahua Chih. México. }\\
		\small{\sl{Universidad Autónoma de Chihuahua}}\\
		\small{\sl{Facultad de Ingeniería}}\\
		
	\end{center}


\begin{abstract}
    \noindent
    In this paper, a convolutional neural network (CNN) was used to classify NMR images of human brains with 4 different types of tumors: meningioma, glioma and pituitary gland tumors. During the training phase of this project, an accuracy of 100\% was obtained, meanwhile, in the evaluation phase the precision was 96\%.
    \keywords{CNN, MRI, Image recognition, Brain tumors, Neural networks.}
    
\end{abstract}

\begin{multicols}{2}

\section{Introduction}

The brain can be defined as a complex organ, located in the skull, that controls the functioning of the nervous system. It is part of the central nervous system and is located in the anterior and upper part of the cranial cavity. Brain functions can be controlled and regulated by most  body and mind functions. This is responsible for vital functions, such as breathing or heart rate regulation, through sleep, hunger or thirst, to higher functions such as reasoning, memory, attention, control emotions and behaviour.

A brain tumor is a mass or group of abnormal cells found in the brain. A tumor that originates for the first time is called a \textit{primary tumor} and is the most common kind of tumor found in the brain. Brain tumors can be classified according to their location, its constituent cells, and whether they are cancerous or benign. The origin of this health issue is still unknown, but it may be related to radiation exposure and aging \cite{presen9}. Although there are many types, in this paper only the following are considered: pituitary gland, meninglioma and glioma.\

In Mexico it is the second and fifth leading cause of death from cancer in the 0-18 and 18-29 age groups, respectively \cite{presen7}. Early diagnosis is crucial to patient survival \cite{presen8}. The most important technique for detecting brain tumors is nuclear magnetic resonance (NMR), a non-invasive and non-radioactive process by which high-resolution images of the inside of the body can be obtained.\

In this work, a convolutional network architecture was used for classification, since it is the best suited for image analysis. A database of 7023 NMR images was used, which contains images of healthy and sick patients. These images were used to train the convolutional network to learn to differentiate between cases of a healthy patient, a patient with glioma, meningioma or pituitary gland.\

This project is a continuation of a previous one \cite{nosotros}, in which we used different machine learning techniques.

\section{Related work}

In  \cite{presen1} a CNN has been used to detect a tumor through MRI. The images were first applied to CNN. A \textit{Softmax} layer was used with a result of 98.67\% of accuracy. Likewise, the precision of the CNN was obtained using the Radial Basis Function (RBF) classifier, achieving a percentage of 97.34\%. It was also used a Decision Tree (DT) classifier, with a result of 94.24\%. In addition to the accuracy criterion, sensitivity, specificity, and precision were used as benchmarks to evaluate the performance of the network.\

In \cite{presen2} MRIs are ordered using CNNs in a public dataset to classify benign tumors and malignant, this to be able to extract characteristics with a better precision. The hybrid model they proposed is a combination of CNN and SVM, however, they were also classified by SVM in order to compare their efficiency. Obtaining the CNN-SVM hybrid with an accuracy of 98.6702\%.

In \cite{presen3} MRI was used to train a hybrid paradigm consisting of \textit{Neural Autoregressive Distribution Estimation (NADE)} and a CNN. This model was later tested with 3064 images with three types of brain tumors. The results obtained show that the NADE-CNN hybrid has a high level classification performance, with an accuracy of 94.49\%.

In \cite{presen4} the authors proposed a CNN method by adding Clustering to extract features, which allows creating a classification into two categories: healthy patient and sick patient. Their dataset has 1892 images, from which 1666 were used for training and 256 for testing. They also performed a preprocessing phase by reducing the images to 227$\times$227 pixels in order to reduce computational times. They compared their proposed method against other types of CNNs with different activation functions, even though they all contained the same structure. The CNN that obtained the highest accuracy had the \textit{Softmax} activation function with accuracy of 98.67\%, and then implemented their Clustering method to the same network, which increased the network's performance up to 99.12\% accuracy.

In \cite{presen5} the authors sought to classify MRI images of people suffering from brain tumors and people without said ailment. They used CNN with a method proposed by themselves, in addition to comparing its accuracy against SVM and DNN. In order to get a higher accuracy during training, the loss layer was added at the end of the training phase to give feedback to the neural network. The accuracies obtained were 83\% for SVM, DNN with 97\% and with their proposed CNN method it was 97.5\%. 

In \cite{presen6} NMR images were classified with ACNN, in their work 253 NMR images were used. The authors transformed the images into gray scale, then changed the size of the images to 256$\times$256 pixels. The dataset was increased to 2912 images with Augmentation, with 50\% healthy patients and 50\% unhealthy patients. His ACCN with Softmax as the activation function achieved an accuracy of 96.7\%.

\section{Methodology}
\subsection{Pre-processing}

The database used during this research project consists of 7023 NMR brain images of different dimensions, in which 5023 images were identified as resonances with the presence of tumors, specifically: 1921 of glioma, 1645 of meningioma and 1757 of pituitary gland, in addition to 2000 images classified as healthy.

Firstly, the images were resized to 200$\times$200 pixels to facilitate their manipulation and processing without losing information. Depending on the format, an image can be divided into layers of red, green and blue hues (RGB image). The images in the dataset were originally in JPEG format, but were changed into PNG because it presented fewer problems during preprocessing. These images were normalized so that the obtained characteristics were in decimal format.
 
Subsequently, the sets of images was divided into 4 folders, according to their category, and stored in another folder, used for the next preprocessing step. With these new images, a Python script was executed, which was able to open directories of each category, plus labeling and processing each image for them to be stored in a PKL file which worked as the dataset of our network. In this part of the process, we also classified the images as: 0 healthy patients, 1 glioma patient, 2 meningioma patients and 3 pituitary gland tumor patients.

80\% of the data was used for training the neural network, leaving the other 20\% for testing, following what has been seen in related works.

\subsection{Network structure}

The structure of the network is shown below

\begin{enumerate}
\item[$\bullet$] First layer: \textit{Convolutional} layer with 32 convolutional filters of 3$\times$3 with \textit{relu} activation function. 

\item[$\bullet$] Second layer: \textit{Maxpooling} layer of 2$\times$2. 

\item[$\bullet$] Third layer : \textit{Convolutional} layer with 64 convolutional filters of 3$\times$3 with \textit{relu} activation function. 

\item[$\bullet$] Fourth layer: \textit{Maxpooling} layer of 2$\times$2.

\item[$\bullet$] Fifth layer: \textit{Convolutional} layer with 128 convolutional filters of 3$\times$3 with \textit{relu} activation function.

\item[$\bullet$] Sixth layer: \textit{Maxpooling} layer of 2$\times$2. 

\item[$\bullet$] Seventh layer: \textit{Convolutional} layer of 128 convolutional filters of 3$\times$3 with a \textit{relu} activation function.

\item[$\bullet$] Eighth layer: \textit{Maxpooling} layer of 2$\times$2.

\item[$\bullet$] Ninth layer: \textit{Flatten} layer.

\item[$\bullet$] Tenth layer: \textit{Dense} layer with 512 neurons with \textit{relu} activation function,

\item[$\bullet$] Eleventh layer: \textit{Dense} layer with 4 neurons with a \textit{Sigmoid} activation function.

\end{enumerate}

\section{Results}

It can be seen in Figure \ref{presición} that the precision had a logarithmic behavior, while the loss (Figure \ref{perdida}) had a negative exponential behavior, which means that the training period was very stable. The model's accuracy during training was 100\%. Furthermore, as it can be seen on the confusion matrix (Figure \ref{matriz}), the results of the evaluation are strongly loaded on the diagonal. During the evaluation, a precision of 96\% was determined.

\begin{center}
\includegraphics[scale=0.4]{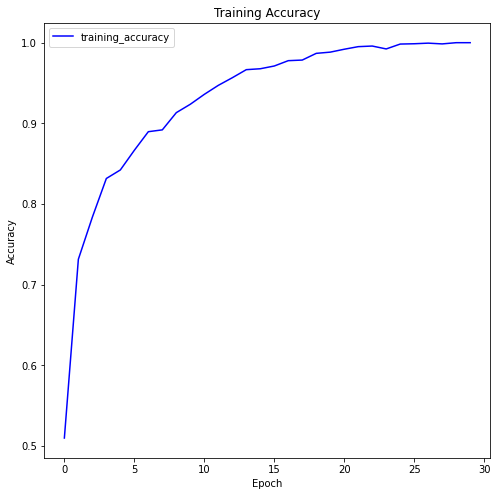}
\captionof{figure}{Accuracy of the model.}
\label{presición}
\end{center}

\begin{center}
\includegraphics[scale=0.4]{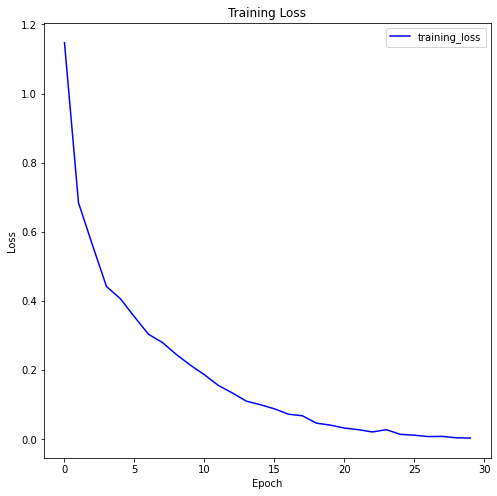}
\captionof{figure}{Loss of the model.}
\label{perdida}
\end{center}

\begin{center}
\includegraphics[scale=0.35]{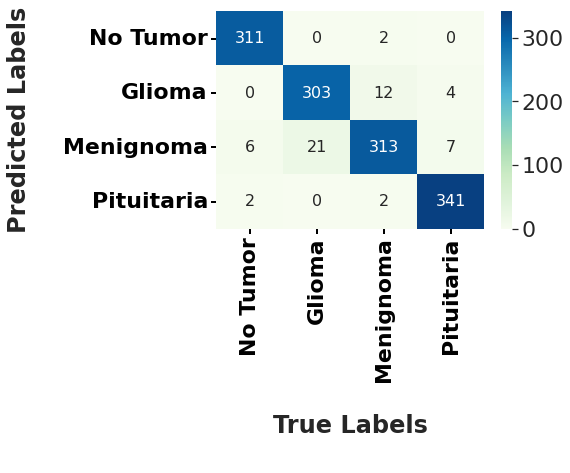}
\captionof{figure}{Confusion matrix of the model.}
\label{matriz}
\end{center}

\section{Conclusions and further work}

An algorithm was developed using convolutional networks that managed to classify NMR images of healthy patients, meningioma, glioma and pituitary gland tumor patients; with an accuracy of 100\% during training phase and a precision of 96\% in the evaluation phase. These results are  consistent on what has been seen in related works. Furthermore, a marked diagonal was observed in the confusion matrix, which indicates that the elaborated model has an acceptable performance and the performance graphs indicate of accuracy and loss doesn't suggest over training. This research can be expanded by including more images to the database and medical advice.


\renewcommand{\refname}{References}

    \nocite{*}
	\bibliographystyle{IEEEtran}
	\bibliography{bib}

\end{multicols}

\end{document}